\documentclass{aa}
\usepackage{graphicx}
\def\approxgt{\mathrel{\hbox{\rlap{\lower.55ex \hbox {$\sim$}}
        \kern-.3em \raise.4ex \hbox{$>$}}}}
\def\approxlt{\mathrel{\hbox{\rlap{\lower.55ex \hbox {$\sim$}}
        \kern-.3em \raise.4ex \hbox{$<$}}}}

\begin{document}
   \title{A hard X-ray view of \\ Giga-Hertz Peaked Spectrum Radio Galaxies}

   \author{M.Guainazzi
          \inst{1},
	  A.Siemiginowska
	  \inst{2},
	  C.Stanghellini
	  \inst{3},
	  P.Grandi
	  \inst{4},
          E.Piconcelli
	  \inst{1},
	  C.Azubike~Ugwoke
	  \inst{5}
          }

   \offprints{M.Guainazzi}

   \institute{$^1$European Space Astronomy Center of ESA, Apartado
              50727, E-28080 Madrid, Spain \\
              \email{Matteo.Guainazzi@sciops.esa.int} \\
	      $^2$Harvard-Smithsonian Center for Astrophysics, 60 Garden St.,
		Cambridge, MA 02138, USA \\
 	      $^3$Istituto di Radioastronomia, I.N.A.F., Via Gobetti 101, I-40129, Bologna, Italy\\
              $^4$Istituto di Astrofisica Spaziale e Fisica Cosmica, I.N.A.F., Via Gobetti 101, I-40129, Bologna, Italy \\
              $^5$Enugu State University of Science and Technology, Nigeria
              }

   \date{Received ; accepted }

   \abstract{ We present the first broadband X-ray observations
of four Giga-Hertz Peaked Spectrum (GPS) radio
galaxies at redshift $\approxlt 1$ performed
by {\it Chandra} and XMM-Newton. These observations more than double the
number of members of this class with
measured spectra in hard ($E > 2$~keV) X-rays.
All sources were detected.
Their radio-to-X-ray spectral energy distributions are
similar, except for PKS~0941-080,
which is X-ray under-luminous by about two orders of magnitude.
The comparison between the full sample of GPS galaxies with measurements in hard
X-rays and a control sample
of radio galaxies rules out intrinsic X-ray weakness as causing a
lower detection rate of GPS sources in X-ray surveys.
Four out of seven GPS galaxies exhibit high X-ray column densities,
whereas for the remaining three this measurement is hampered by
the poor spectral statistics.
Bearing in mind the low number statistics in
both the GPS and the control sample, the average column density
measured in GPS galaxies is larger than in
FR~I or Broad Line Region FR~II radio galaxies, but consistent
with that measured in High-Excitation FR~II galaxies. This
leads to a location the absorbing gas in an obscuring ``torus'',
which prevents us from observing the nuclear region along lines-of-sight
perpendicular to the radio axis. This interpretation is
supported by the discovery of rapid (timescale $\sim 10^3$~s)
X-ray variability in the GPS galaxy
COINSJ0029+3456, and by an almost
order-of-magnitude difference between the HI column density
measured in radio and X-rays in PKS0500+019.
            }

\authorrunning{Guainazzi et al.}

\titlerunning{Hard X-ray view of GPS galaxies}

   \maketitle
%

\section{Introduction}

GHz Peaked Spectrum (GPS) radio
sources are extremely powerful $(L_{radio} \approxgt 10^{45}$
erg sec$^{-1}$), very compact (10--100~mas, 10--1000~parsecs)
sources characterized
by a simple convex radio spectrum peaking
near 1 GHz
(O'Dea 1998 and references therein). 
GPS sources make
up about 10\% of the sources selected at
frequencies around 5~GHz. They
often exhibit symmetric structures on the parsec scale,
reminiscent of those present in extended radio galaxies
on much larger scales.
About half
of the GPS sources are classified as {\it ``GPS galaxies"}.

GPS sources - together with 
Compact Steep Spectrum (CSS) sources
(\cite{odea98}) - may play an important role in the
history and evolution
of the radio universe. It was originally suggested that
they may represent
radio galaxies in the early stage of their life
(typical ages $< 10^4$~years;
Phillips \& Mutel 1982; Carvalho 1985;
De Young 1993; Fanti et al. 1995; Readhead et al
1996; Murgia 2003). This possibility was recently
supported by the detection of
micro arc-seconds hotspot proper motions in a sample of $\simeq$10 GPS
galaxies (Poladitis \& Conway 2003).
However, evidence for weak
$\sim$10~kpc scale radio emission
in 5--10\% of all known GPS (Stanghellini et al. 1998
and references therein)
suggested that the radio
activity may be either recurrent
(Baum et al. 1990), or smothered by the inflow of
``fresh'' matter after merging or interaction with a gas
rich companion. Alternatively, as originally suggested 
by Gopal-Krishma \& Wiita (1991), GPS
sources could remain
compact during their radiative lifetime
because interaction with dense
circumnuclear matter impedes
their full growth.
Semi-analytical
models (Carvalho 1994;
1998) and hydrodynamic simulations (De Young 1993)
have indeed shown that InterStellar Medium (ISM) average
densities in
the range 1--10~cm$^{-3}$ could be very efficient
in preventing the development of large scale radio
structures.

Measurements in X-rays could provide fundamental clues on the
nature of GPS sources and
consequently on the
origin and evolution of the radio power in the universe.
A hot ($T \sim 10^{6-7}$~K) and tenuous ISM phase
with
$n \sim 1$~cm$^{-3}$ should be a copious source
of soft X-rays.
However,
the recent Chandra discovery of the X-ray cluster associated with the CSS
quasar 3C~186 rules out the confinement of the radio source by external medium
in this quasar (\cite{siemiginowska05a}).
Alternatively, a distribution of cold clouds would
imprint an X-ray photoelectric
cutoff on the soft X-ray spectrum.
It has been noted that GPS sources
are rather elusive in X-rays.
In the ROSAT All Sky Survey,
GPS/CSS quasars
exhibited a detection rate three times lower than
radio-loud quasars of comparable power
(\cite{baker95}).
Whether this low detection rate is due to intrinsic
weakness or to obscuration of the active nucleus
is still largely unknown. Prior to the launch
of {\it Chandra} and XMM-Newton, high-quality
CCD-resolution hard ($E > 2$~keV)
X-ray spectra were available
for two GPS galaxies only: NGC~1052
(\cite{guainazzi00})\footnote{see Sect.~5.2
for a discussion on the GPS classification of
this radio galaxy} and 4C+12.50 (\cite{odea00}).
Both exhibit a high
X-ray obscuring column density ($N_H > 10^{22}$~cm$^{-2}$).

In order to gain insight into the X-ray under-luminosity of
GPS galaxies , we have performed {\it Chandra} and
XMM-Newton
observations of 5 GPS galaxies at
$z \approxlt 1$. This paper presents the
results of this program. The observations discussed
here represent the first hard X-ray measurements
of the target sources. They more than double the
number of GPS galaxies with hard X-ray measurements.
The results of the observation of Mkn~668 (OQ+208),
performed in the framework of our program, were
presented by Guainazzi et al. (2004).
This observation allowed
the discovery of the first radio-loud Compton-thick
($N_H > \sigma_t^{-1}$) AGN.

Unless otherwise specified:
energies are quoted in the source
reference frame; uncertainties on the spectral
parameters are at the 90\% confidence level
for one interesting parameter; upper limits are
also at the 90\% confidence level;
other uncertainties are at the 1$\sigma$ level;
in the
calculation of the luminosities, we adopted
a Hubble constant of 70~km~s$^{-1}$~Mpc$^{-1}$
(\cite{bennett03}); data points in the spectral plots
correspond to a signal-to-noise $>3$.
We have used throughout the paper the
Gehrels algorithm (\cite{gehrels86}) to
estimate the statistical uncertainties associated
with Poissonian-distributed measurements.

\section{The sample}

In this section we present a short description of the
radio properties of the sources discussed
in this paper, supporting their
classification as ``GPS''
galaxies. We require
{\it both} a convex radio spectrum {\it and} a symmetric
radio morphology to classify a radio galaxy as
GPS.
\\[0.25cm]
\noindent
{\it COINS~J0029+3456} (B~0026+346) is a galaxy
with $m_r$=21.0\footnote{$m_r$ and $m_i$
are the magnitudes in the $r$ and $i$
band, respectively, according to the Gunn system.}
at z=0.517 (\cite{snellen96,zensus02}). It
has a broad radio spectrum, which
is rather flat at high frequencies (\cite{kuhr81}).
It exhibits a compact symmetric morphology with an overall
size of 35~mas (190~pc).
\\[0.1cm]
\noindent
{\it PKS~0500+019} (J0503+0203) is a radio galaxy
with $m_i$=21.0 at z=0.583 (\cite{carilli98}) with
a Compact Symmetric Object (CSO)
morphology which extends for about 15~mas (84~pc).
Stickel et al. (1996)
reported an emission line at a higher redshift,
therefore they consider the GPS radio source
associated with a background quasar and the galaxy as an intervening object.
De Vries et al. (2000), however, did not find any trace of
this emission line in their optical spectra.
\\[0.1cm]
\noindent
{\it PKS~0941-080} is a radio galaxy at $z=0.228$
with a clear convex radio spectrum and a rather clear CSO morphology
(\cite{dallacasa97}).
\\[0.1cm]
\noindent
{\it PKS~2128+048} (J2130+0502)
is a galaxy with $m_r$=23.3 at z=0.99 (\cite{snellen96}) with a
clear GPS radio spectrum and a rather clear CSO morphology of 35~mas
(220~pc) (\cite{stanghellini97}, 2001).

Table~\ref{tab1} reports the list of observations
discussed in this paper.
\begin{table*}
\caption{Log of the observations discussed in this paper.
``Threshold'' indicates the maximum background count rate used in the
generation of appropriate time intervals for the extraction of spectra.
The ``Radius'' refers to the source
extraction region}
\begin{center}
\begin{tabular}{lcccccc} \hline \hline
Source & z & $N_{H,Gal}$ & Start date & T$_{exp}$$^a$ &  Threshold (s$^{-1}$) & Radius ($\arcsec$)$^a$ \\
& & (10$^{20}$~cm$^{-2}$) & &  (ks) &  \\ \hline
COINS~J0029+3456 & 0.571 & 5.6 & 08-Jan-2004 & 12.2/10.8 & 0.25/1.0 & 35/40 \\
PKS~0500+019 (1) & 0.584 & 8.3 & 09-Mar-2004 & 2.0/1.9 & 2.0/1.0 & 20/30 \\
PKS~0500+019 (2) & & & 18-Aug-2004 & 9.6/7.2 & 0.0/0.0 & 20/30 \\
PKS~0941-080  & 0.228 & 3.7 &  26-Mar-2002 &  5.3 & ... &  1.4 \\
PKS~2128+048  & 0.990 & 5.2    &  11-Oct-2002 &  5.7 & ... & 2.4 \\
\hline \hline
\end{tabular}
\end{center}
\noindent
$^a$for XMM-Newton MOS/pn and {\it Chandra} ACIS, respectively
\label{tab1}
\end{table*}
The March 2004 observation of PKS0500+019 was heavily affected by high-particle
background, which reduced the usable exposure time to $\simeq$2~ks. The target
was therefore successfully re-observed in August.

\section{XMM-Newton results}

The XMM-Newton data 
were reduced with SAS v6.0
(\cite{gabriel03}), using the most updated calibration
files (January 2005). In this paper, only data from the
EPIC cameras (MOS; Turner et al. 2001; pn,
Str\"uder et al. 2001) will be discussed.
Event lists from the two MOS cameras were
merged before accumulation of any scientific
products. Single to double (quadruple)
events were used to accumulate pn (MOS) spectra.
High-background particle flares were removed by
applying fixed thresholds on the single-event,
$E>10$~keV, $\Delta t = 10$~s light curves.
These thresholds, as well as the radius of the
source scientific product circular extraction regions,
were optimized to maximize the signal-to-noise
ratio, and are listed in Table~\ref{tab1}.
Background scientific
products were extracted from annuli around the source
for the MOS, and from
circular regions in the same chip
for the pn, at the same height in detector coordinate as the
source location.
Spectra were binned
in order to oversample the intrinsic
instrumental energy resolution by a factor
$\ge$3, and to have at least 25 counts in each
background-subtracted spectral channel. This
ensures that the $\chi^2$ statistics can
be used to evaluate the quality of the
spectral fitting. pn (MOS) spectra were
fitted in the 0.35--15~keV (0.5--10~keV)
spectral range.

\subsection{Source identification}

In Table~\ref{tab3} we list the coordinates of the nominal
\begin{table}
\caption{GPS sample sources X-ray coordinates. ``$d$'' indicates the
distance between the X-ray and the radio coordinates, the latter
extracted from the NED catalogue. ``CTS'' are net counts in the
0.5--10~keV and 0.1--11~keV energy bands for the XMM-Newton EPIC
and the {\it Chandra} ACIS, respectively. The values for PKS~0500+019
refer to the second observation only}
\begin{center}
\begin{tabular}{lcccc} \hline \hline
Source & RA & Dec & d ($\arcsec$) & CTS \\
& (J2000) & (J2000) & &  \\ \hline
0029 & 00$^{\rm h}$29$^{\rm m}$14$^{\rm s}$.1 & +34$^{\circ}$56$\arcmin$32$\arcsec$ & 1.7 & $1030 \pm 40$\\
0500 & 05$^{\rm h}$03$^{\rm m}$20$^{\rm s}$.9 & +02$^{\circ}$03$\arcmin$05$\arcsec$ & 4.5 & $1350 \pm 40$ \\
0941 & 09$^{\rm h}$43$^{\rm m}$36$^{\rm s}$.9$^a$ & -08$^{\circ}$19$\arcmin$31$\arcsec$$^a$ &  ... & $10 \pm 3$ \\
2128 & 21$^{\rm h}$30$^{\rm m}$32$^{\rm s}$.8 & +05$^{\circ}$02$\arcmin$08$\arcsec$ & 1.1 &  $92 \pm 10$ \\
\hline \hline
\end{tabular}
\end{center}

\noindent
$^a$fixed to the radio position
\label{tab3}
\end{table}
targets in our XMM-Newton observations. They were
determined on the 0.5--10~keV image extracted from
merged event lists of the three EPIC cameras.
Typical statistical errors are
$\simeq$1$\arcsec$, whereas typical systematic uncertainties
on the absolute attitude reconstruction are
$\simeq$1.5$\arcsec$. For PKS0500+019 (2$^{\rm nd}$ observation)
the X-ray
position is shifted by about 4.5$\arcsec$ in RA with respect to the
position of the radio source (\cite{condon98,ma98}),
corresponding to about 25~kpc at the source redshift.
This difference exceeds the expected measurement uncertainties.
The XMM-Newton position agrees well with that of the
closest source in the ROSAT PSPC catalog (\cite{voges00}).
No optical counterpart exists for either the X-ray or
the the radio position.
The paucity of X-ray/optical coincidences 
in the XMM-Newton field prevented us from performing an
independent calibration of the absolute astrometry in
this observation. Nonetheless, a comparable shift in RA
is observed between several XMM-Newton sources and the
closest Digital Sky Survey plate source. Although this
evidence does not
constitute proof, we will consider in the
following such a shift as due to residual inaccuracies in
the attitude reconstruction.

\subsection{Timing analysis in COINS~J0029+3456}

\begin{figure}
   \centering
   \includegraphics[width=7cm, angle=-90]{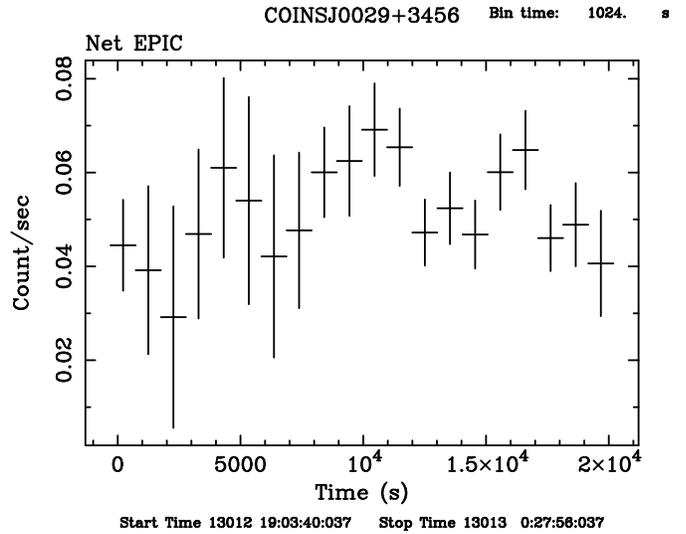}
\caption{0.5--10~keV
EPIC (pn {\it and} MOS) light curve during
the XMM-Newton observation of COINS~J0029+3456.
The binning time is $\Delta t=$~1024~s. Times are in seconds after
the observation start.
              }
\label{fig6}
\end{figure}

The light curves observed during the XMM-Newton observations
are consistent with being constant
in all energy ranges, except for COINS~J0029+3456, where
variability of a factor of $\simeq$3 within a few thousand
seconds was observed (Fig.~\ref{fig6}). By fitting
consecutive segments of the light curve
with a linear function, we estimate a
typical luminosity change rate of
$(5.0 \pm 2.5) \times 10^{43}$~erg~s$^{-1}$~hour$^{-1}$.
We find only marginal evidence that the variability
is primarily due to hard X-rays. The reduced $\chi^2$
resulting from fitting with a constant line the
$\Delta t = 1024$~s binned light curves is 0.86
and 1.27 in the 0.2--1 and 1--10~keV energy band, respectively.
Consequently, no significant
variability of the hardness ratio
is associated with these flux changes. We will
therefore discuss the time-averaged spectrum of this source
in Sect.~3.3. COINS~J0029+3456 also shows a remarkable
historical variability in the soft X-ray band. The 1~keV
flux density decreased from $100 \pm 30$~nJy (Einstein/IPC, 1980)
and $63 \pm 15$~nJy (ROSAT/PSPC, 1992) (\cite{kollgaard95};
our re-analysis of archival data) to
$29 \pm^9_7$~nJy as measured about 10
years later (January 2004)
by XMM-Newton/EPIC.
It is impossible to tell whether
a change of the column density of the intervening absorber
contributes to this variability,
due to the lack of statistical quality and energy resolution
of the measurements prior to XMM-Newton.

\subsection{Spectral analysis}

Spectra of the EPIC cameras have been simultaneously fitted, allowing
only a free cross-normalization factor between them.
We have applied a ``baseline model'', constituted by a power-law
modified by photoelectric absorption. We have modeled the absorption
simultaneously with
two components: a) the intervening gas along the
line of sight in our Galaxy, with a column density $N_{H,Gal}$ held
fixed to the value determined by the 21~cm radio maps (\cite{dickey90})
and;
b) a ``local'' absorber located at the same redshift as the systemic
velocity of each galaxy; its column density, $N_H$, was left as
a free parameter in the fit. All the continuum
parameters were left free in the fit as well. This baseline
model yields a good fit for all
observations.
The spectral parameters derived from the two observations of
PKS~0500+019 are formally consistent within the statistical uncertainties,
except for a $\simeq$40\% larger average flux in the later one.
We will consider in the following the results obtained during
the second observation of PKS~0500+019,
as their associated statistical uncertainties are
smaller.

\begin{figure*}[hbt]
   \centering
\hbox{
   \includegraphics[width=6cm, angle=-90]{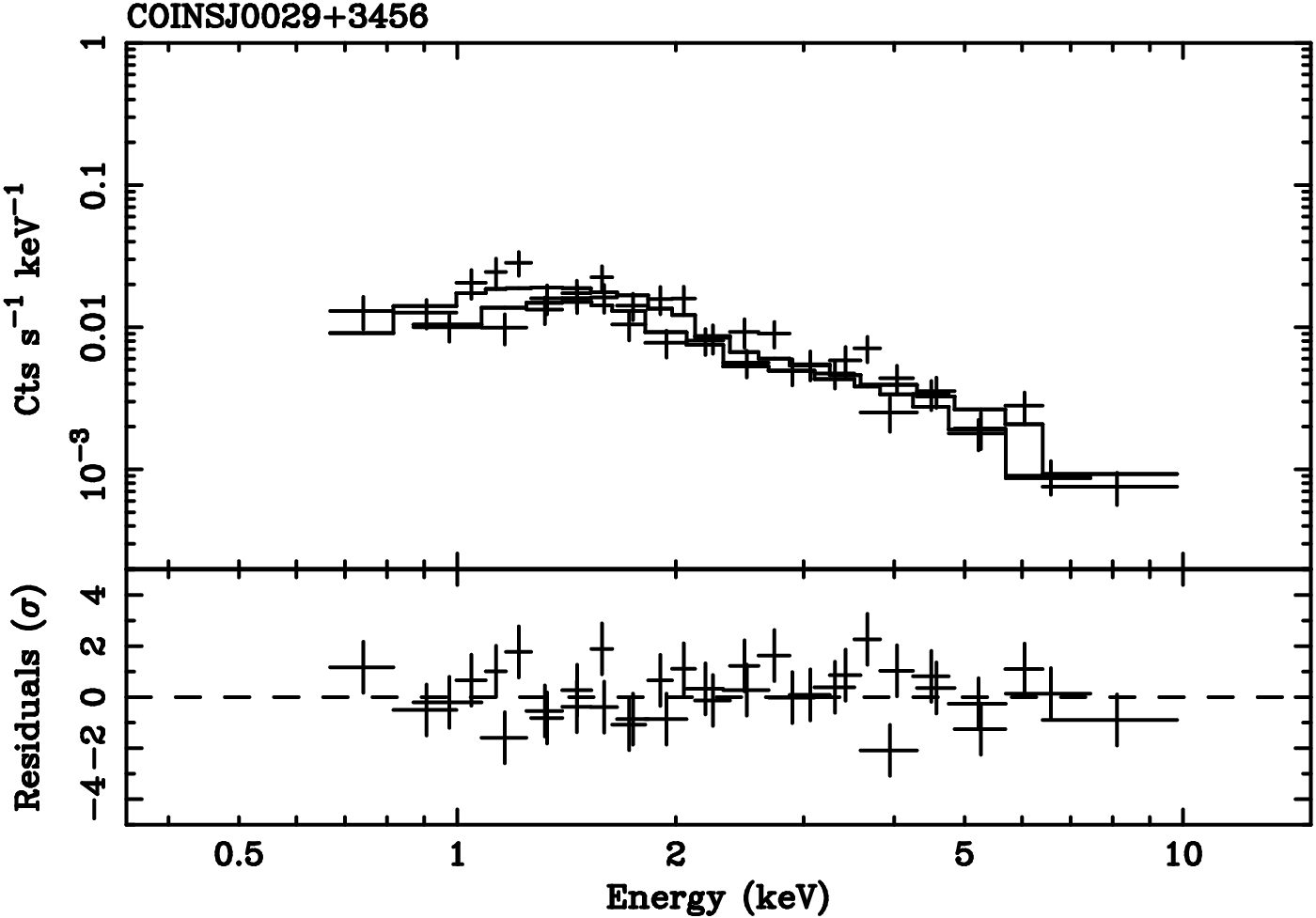}
   \hspace{0.5cm}
   \includegraphics[width=6cm, angle=-90]{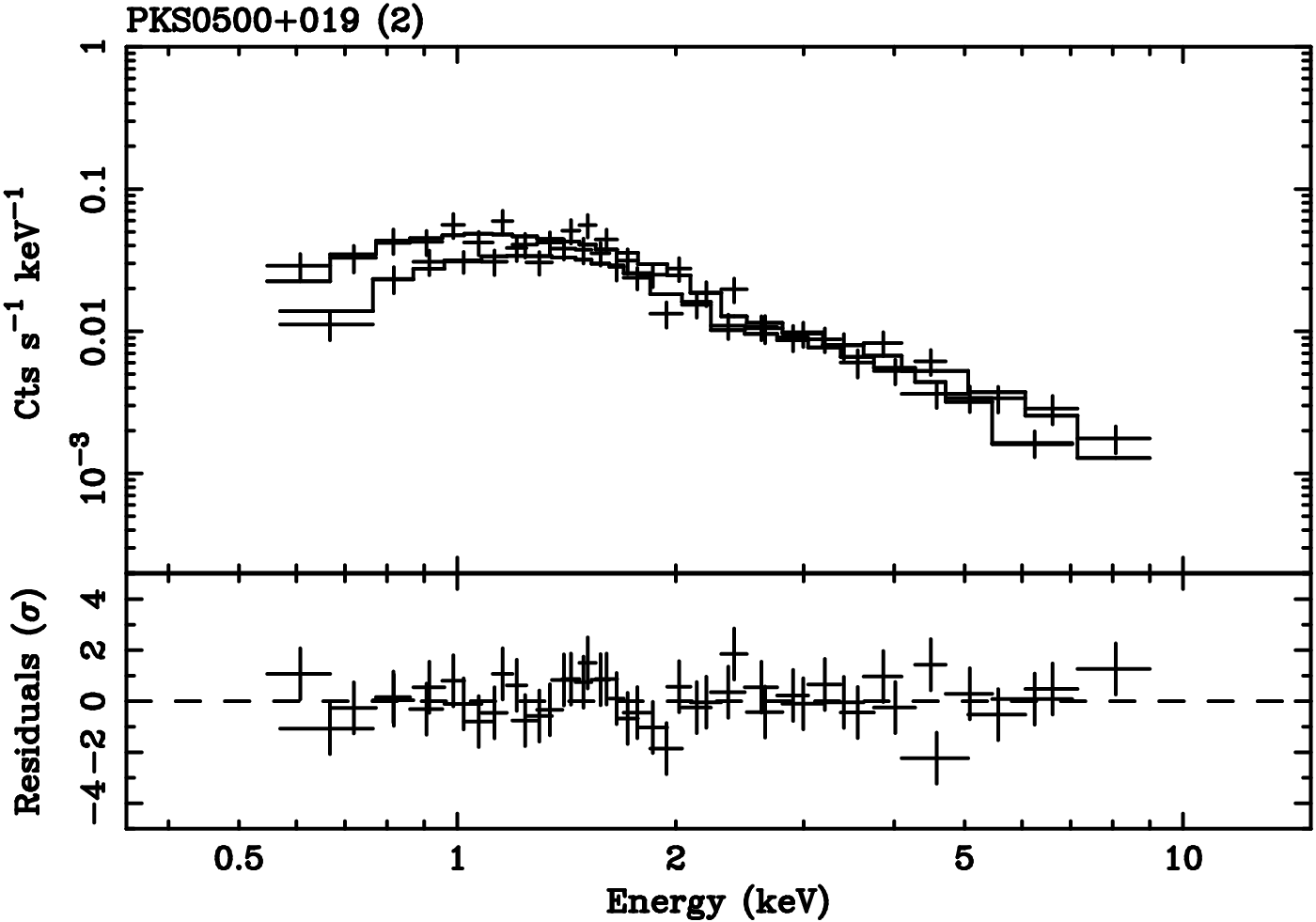}
}
\caption{Spectra ({\it upper panels}) and
residuals in units of standard deviations
({\it lower panels}) when the best fit model of
Table~\ref{tab2} is applied to the XMM-Newton/EPIC
spectra of COINS~J0029+3456 and PKS~0500+019.
              }
\label{fig2}
\end{figure*}

Table~\ref{tab2} summarizes the spectral parameters of the
XMM-Newton GPS sample presented in this paper.
\begin{table*}
\caption{Spectral results on the XMM-Newton GPS sample described in this paper.
The model employed to fit the data is a photoelectrically absorbed
power-law.
The obscuring gas with column density $N_H$ is assumed to be located
at the systemic velocity of the galaxy.
$\Gamma$ is the photon index of the power-law, $F$ the observed
flux, $L$ the {\it intrinsic} luminosity, corrected for absorption.}
\begin{center}
\begin{tabular}{lccccc} \hline \hline
Source & $N_H$ & $\Gamma$ & $F^a$ & $L^b$ & $\chi^2/\nu$ \\
& (10$^{21}$~cm$^{-2}$) & &  & & \\ \hline
COINSJ~0029+3456 & $10 \pm^5_4$ & $1.43 \pm^{0.20}_{0.19}$ & 2.7 & $2.3 \pm 0.2$ & 38.9/34 \\
PKS~0500+019 (1) & $10 \pm^{11}_9$ & $2.0 \pm^{0.7}_{0.5}$ & 3.9 & $3.8 \pm 0.9$ & 10.3/9 \\
PKS~0500+019 (2) & $5 \pm^3_2$ & $1.62 \pm^{0.21}_{0.19}$ & 5.6 & $5.0 \pm 0.6$ & 33.1/44 \\
PKS~0941-080 & ... & 2$^c$ & 0.07 & $0.009 \pm 0.008$ & ...$^d$ \\
PKS~2128+048 & $3.0 \pm^{8.1}_{3.0}$ & $1.5 \pm^{0.6}_{0.7}$ & 1.6 & $4.4 \pm 1.1$ & 1.7/4 \\
\hline \hline
\end{tabular}
\end{center}

\noindent
$^a$in units of 10$^{-13}$~erg~cm$^{-2}$~s$^{-1}$ in the 0.5--10~keV energy band (observer's frame)

\noindent
$^b$in units of 10$^{44}$~erg~s$^{-1}$ in the 2--10~keV band (source frame), corrected for absorption

\noindent
$^c$fixed

\noindent
$^d$not enough independent channels for a spectral fit to be possible

\label{tab2}
\end{table*}
The spectra and residuals against the best fit models
are shown in Fig.~\ref{fig2}.

\section{Chandra results}

Two GPS galaxies, PKS~0941-080 and PKS~2128+048, were observed with the
{\it Chandra} Advanced CCD Imaging Spectrometer (ACIS-S,
\cite{weisskopf02}) as part of a program
to observe a sample of GPS sources
(Siemiginowska et al., 2005b). The short 5~ksec observations were
mainly
scheduled to measure the X-ray flux.  The sources were located $\sim
35\arcsec$ from the default aim-point position (to avoid node
boundaries) on the ACIS-S backside illuminated chip S3 (Proposer's
Observatory Guide
\footnote{http://asc.harvard.edu/proposer/POG/index.html}). The
1/8 sub-array CCD readout mode of one CCD only was used resulting in
a 0.441~sec frame readout time.

The X-ray data analysis was performed in CIAO
3.1\footnote{http://cxc.harvard.edu/ciao/} with the calibration files
from the CALDB 2.28 data base. Note that the ACIS-S contamination file
{\tt acisD1999-08-13contamN0003.fits} which incorporates effects of
the ACIS-S contamination into the effective area was included in our
analysis. The spectra were rebinned according to the same
criteria employed for XMM-Newton (see Sect.~3).

The two galaxies were detected by Chandra with the count rates listed
in Table~\ref{tab3}. For PKS~0941-080 the total number of net counts
($\simeq$10) is too low to perform a true spectral fitting.
In PKS~2128+048 a simple power-law model with photoelectric
absorption yields a good fit. A column
density of $N_H \simeq 3 \times 10^{21}$~cm$^{-2}$ is
measured. However this measurement is still consistent with
0 if the large statistical uncertainties are taken into account.

\section{Discussion}

\subsection{Spectral Energy Distributions}

\begin{figure*}[hbt]
   \centering
   \includegraphics[width=12cm, angle=90]{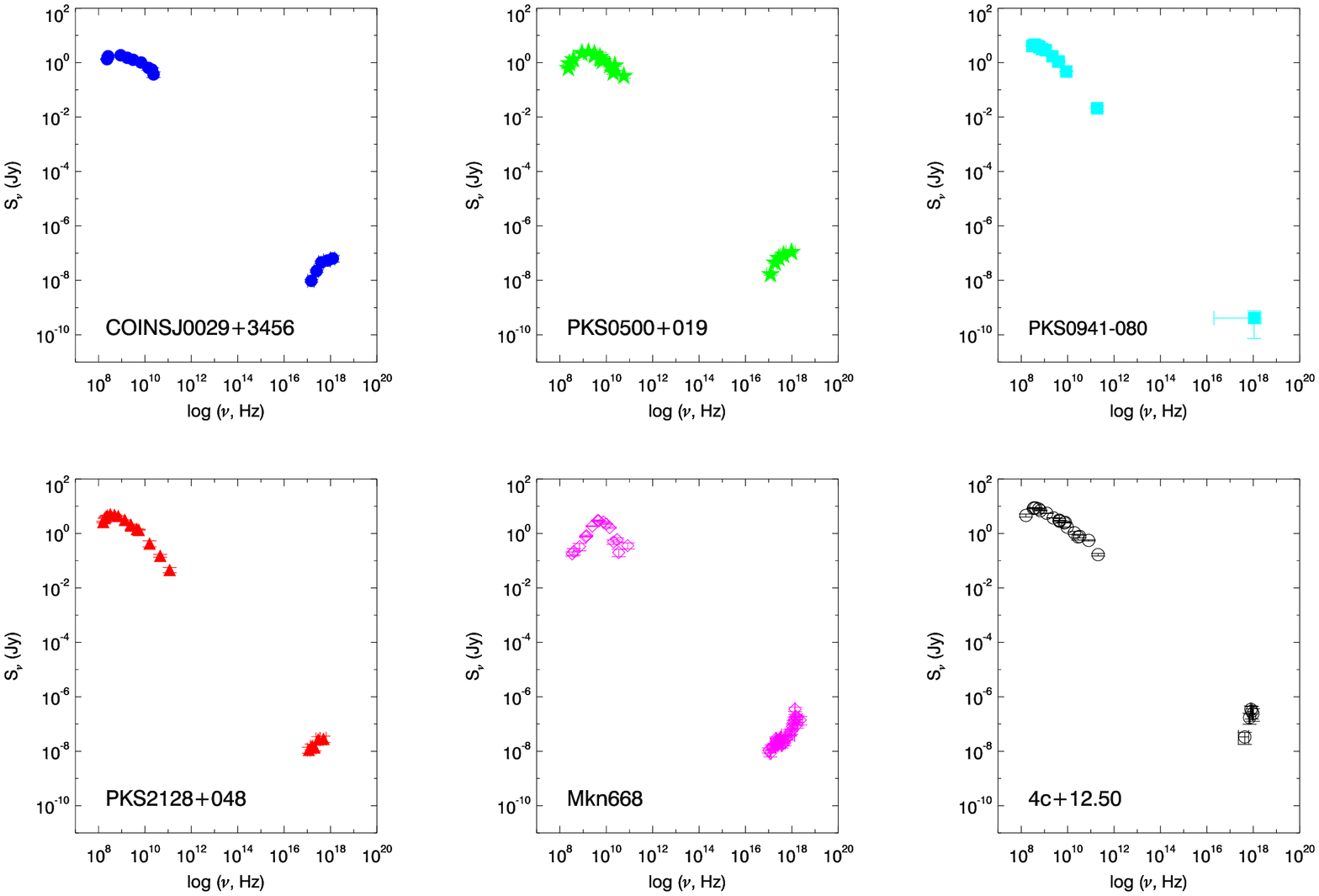}
\caption{Radio-to-X-ray Spectral Energy Distributions
for the objects of our sample, 4C+12.50 (\cite{odea00})
and Mkn~668
(\cite{guainazzi04}). Radio measurements
represent not simultaneous observations
compiled from the NASA Extragalactic Database
(NED).}
\label{fig7}
\end{figure*}

\begin{figure}
   \centering
\hbox{
   \hspace{-0.5cm}
   \includegraphics[width=6.25cm, angle=90]{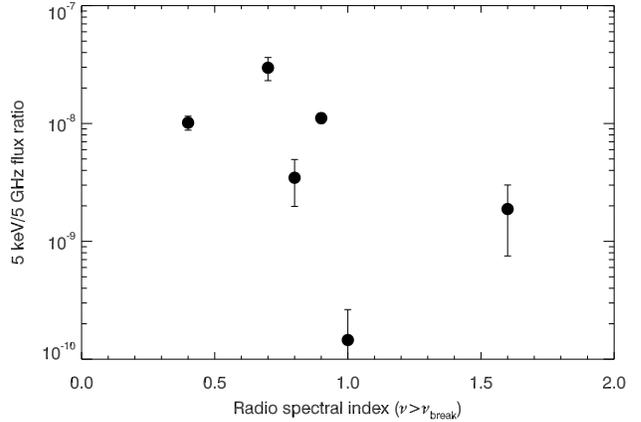}
}
\caption{5~keV to 5~GHz flux density ratio as
a function of the radio spectral index
above the break frequency. The slope
of a linear least squares fit is
$(-1.2 \pm 0.6) \times 10^{-8}$. The
slope
does not significantly change if the
data point corresponding to PKS~0941+080
is excluded from the fit.}
\label{fig10}
\end{figure}

In Fig.~\ref{fig7} we show the radio-to-X-ray
Spectral Energy Distributions (SEDs) for the
GPS galaxies discussed in this paper, as well
as for 4C+12.50 (\cite{odea00})
and Mkn~668 (\cite{guainazzi04}).
The 5~GHz to
5~keV flux density ratios of 5 out of
6 objects are within a factor of 15. On
the other hand, in
PKS~0941+080 the 5~GHz to
5~keV flux density ratio is
about two orders-of-magnitude lower than the
average for the rest of the sample. Whether this is due to
intrinsic weakness or extreme ({\it i.e.} Compton-thick)
obscuration is impossible to tell with the current
data, and would require very deep
X-ray exposures to be finally
elucidated. There is no correlation
between the X-ray to radio flux density ratio
and the radio break frequency; however,
an anti-correlation at the 2$\sigma$ level
is found between the X-ray to radio flux density ratio
and the radio spectral index above the
frequency break (\cite{stanghellini98};
see Fig.~\ref{fig10}). 

Mkn~668 is a peculiar object in our
small sample. Not only does it exhibit the
largest measured obscuration among the GPS
galaxies for which hard X-ray measurements
are available (being one of the very few known
radio-loud Compton-thick AGN),
it also shows a prominent soft X-ray
excess above the extrapolation of the obscured
primary AGN continuum. None of the other GPS
exhibits this component in the
XMM-Newton or {\it Chandra} spectra. Its origin is still to be
fully elucidated, the most
likely possibilities being scattering of the
AGN radiation or inverse-Compton recoil of the
large far infrared emission
(\cite{mazzarella91,guainazzi04}).

\subsection{Comparison with a control sample of
``normal'' radio galaxies}

In this Section we compare the spectral properties of
the full sample of GPS galaxies for which hard X-ray measurements
are available with a
control sample of Radio Galaxies (RGs).
The former includes, with the objects discussed in
this paper: 4C+12.50
(\cite{odea00}),
4C+55.16 (\cite{iwasawa99}),
OQ+208 (\cite{guainazzi04}). We do {\it not}
include the radio galaxy NGC~1052, despite it
being sometimes classified as a ``recurrent'' GPS
source (\cite{tingay03,edwards04}).
NGC~1052 has a 2-sided jet morphology at the
mas scale, without evidence of micro hot spots
(\cite{kadler04,satyapal04}).
The hot spots are instead present at large scales,
implying that the the jets are a continuous
structure, which ends at the kpc scale.

The control sample is taken from
a compilation of radio-loud AGN X-ray spectra
discussed by
Sambruna et  al. (ASCA, 1999; S99 hereafter),
Donato et al. ({\it Chandra}, 2004; D04), and
Fiocchi et al. (BeppoSAX, 2005; F05).
S99 provides in the redshift range of the GPS sample
a not statistically complete sample
of 15 radio galaxies of mixed FR~I and FR~II types,
and optical (broad/narrow line) properties.
To this, F05 add five more
FR~II and D04 three more FR~I galaxies, again of mixed optical types.
This comparison aims
to address the following basic questions:
a) are GPS galaxies intrinsically X-ray weaker than ``normal'' radio galaxies?;
b) are GPS galaxies more obscured than ``normal'' radio galaxies?

\subsubsection{Are GPS galaxies intrinsically X-ray weaker than ``normal'' radio galaxies?}

\begin{figure}
   \centering
\hbox{
   \hspace{-1.0cm}
   \includegraphics[width=9.5cm]{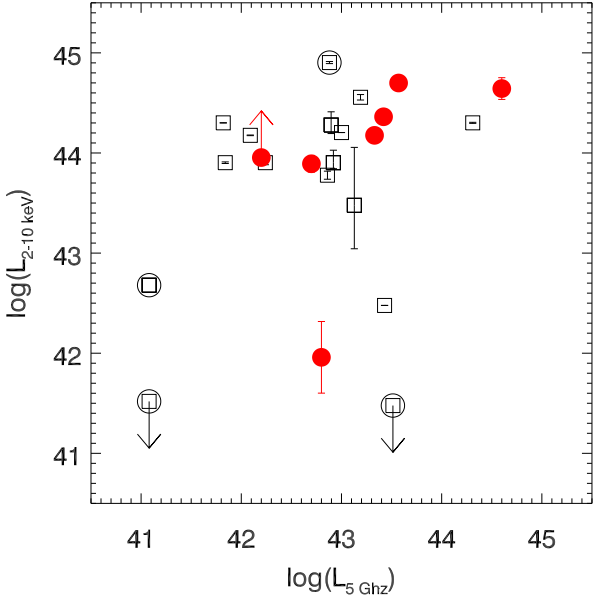}
}
\caption{5~GHz versus 2--10~keV for the
control sample radio galaxies ({\it empty squares})
and the GPS galaxies sample ({\it filled circles}).
{\it Empty circles} surround the FR~I radio
galaxies data points.
              }
\label{fig3}
\end{figure}

{\it No}. In Fig.~\ref{fig3} we show the location of the GPS sample
and of the control radio galaxy sample in the 5~GHz versus 2--10~keV
luminosity plane (we stress that the X-ray luminosity is {\it
intrinsic}, {\it i.e.} corrected for absorption).
GPS galaxies occupy the same region of the $L_X$ versus $L_{radio}$
plane
as ``normal'' radio galaxies across the whole range of X-ray luminosities
covered by the sample. The
average luminosity ratios
$R_{Xr}$ are: $\log [R_{Xr}(GPS)] = 0.7 \pm 0.3$,
$\log [R_{Xr}(RG)] = 0.69 \pm 0.17$.

The origin of the X-ray continuum in GPS galaxies
is still largely unknown. Observed spectral
indices are in the range 1.4--2.0, therefore consistent
with both accretion- (\cite{perola02,risaliti02}) or
beamed-dominated (\cite{fossati98})
systems. The upper limits on the equivalent
width of the K$_{\alpha}$ fluorescent
iron line ($\approxlt$150~eV) are not
stringent enough to rule out the former
explanation.
Moreover, for none of the objects
presented in this paper are measurements at 7000$\AA$
available, which may contribute to elucidate
the nature of the central core continuum
through the radio-to-UV correlation in
Chiaberge et al. (1999).

\subsubsection{Are GPS galaxies more obscured than ``normal'' radio galaxies?}

\begin{figure*}[hbt]
   \centering
\hbox{
   \includegraphics[width=9cm]{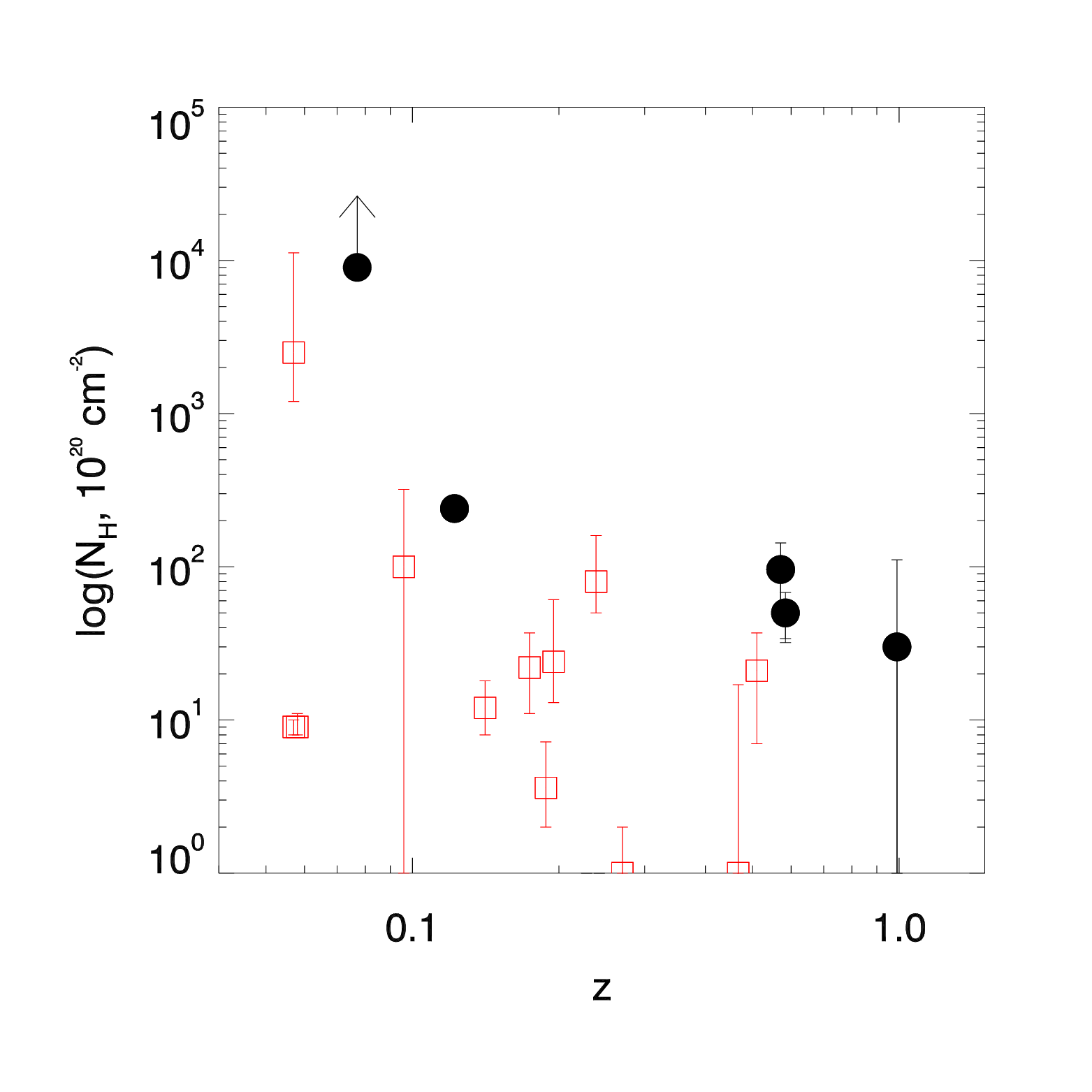}
   \includegraphics[width=9cm]{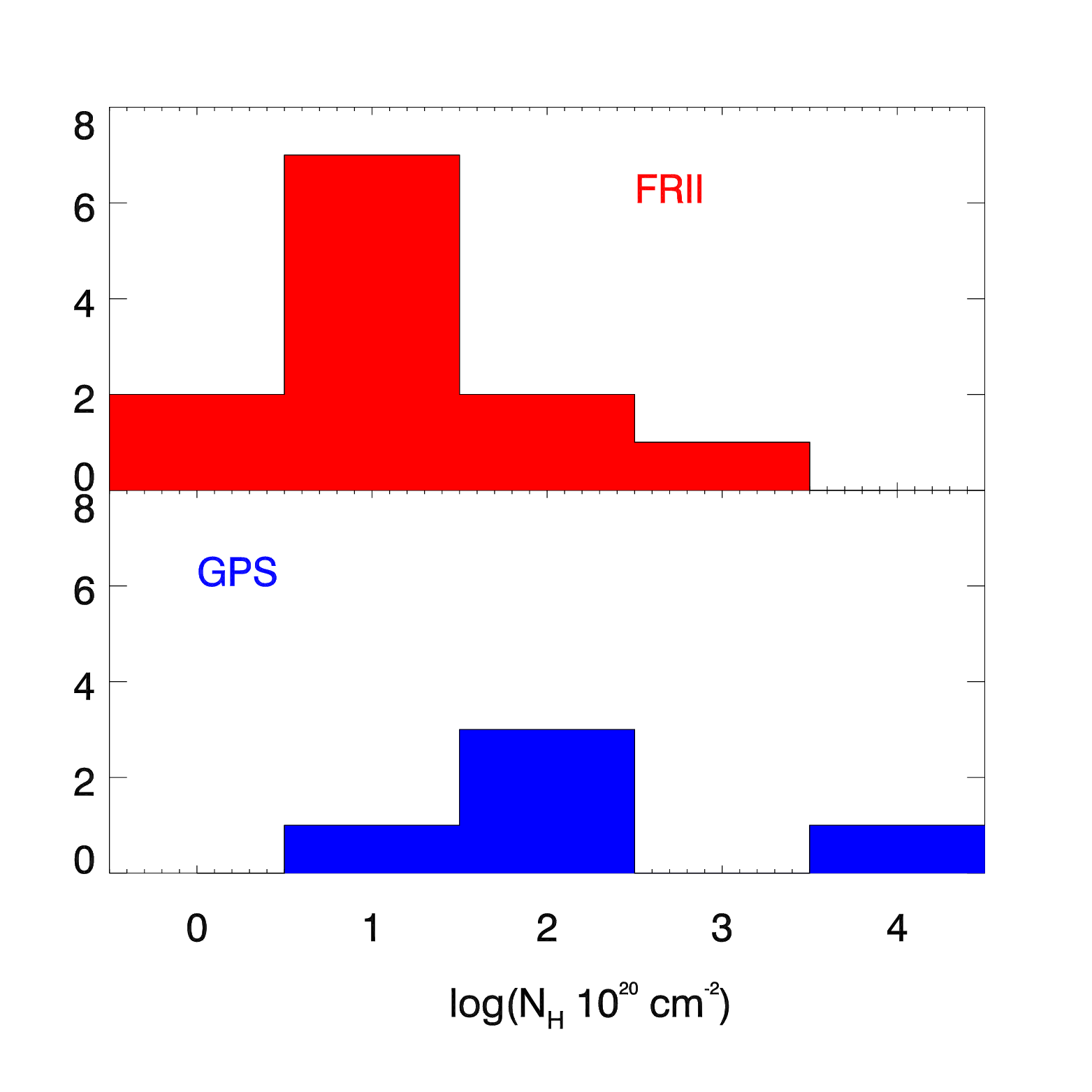}
}
\caption{{\it Left panel}: $N_H$ versus $z$ distributions
for GPS galaxies
({\it filled circles}) and the control sample of FR~II radio
galaxies ({\it empty squares}). {\it Right panel}:
the $N_H$ distribution functions for
FR~II radio galaxies ({\it top}) and GPS galaxies.
              }
\label{fig4}
\end{figure*}

This may be true.
In a recent {\it Chandra} survey of FR~I RGs (D04), only one out
of 15 objects exhibit a column density $\simeq$10$^{22}$~cm$^{-2}$,
whereas for 13 of them $N_H \le 10^{21.3}$~cm$^{-2}$ (D04).
In Fig.~\ref{fig4} we show the measured column
densities for the GPS sample and the FR~II sample
of S99 and F05, together with the $N_H$ distribution function
for the two classes. The apparent anti-correlation between
$N_H$ and $z$ is most likely due to a selection effect,
as farther objects with greater obscuration are
more easily missed. All the measurements of $N_H$ in
GPS galaxies lie on the upper envelope of the plot.
With the caveat of the small numbers, the two
distributions are different at the 98.3\% level,
according to the Kolmogorov-Smirnov (K-S) test. The
difference is primarily due to the relative lack of
unabsorbed GPS galaxies. Only one out of five GPS
galaxies ($< 46\%$) has $N_H \approxlt 10^{22}$~cm$^{-2}$,
compared to $75 \pm 26\%$ RGs. If the X-ray weakness of
PKS~0941-080 is due to high obscuration,
or we include in the sample NGC~1052
($N_H \simeq 2 \times 10^{23}$~cm$^{-2}$,
\cite{guainazzi00})
the evidence for greater absorption in
GPS galaxy is even more significant
(see as well Guainazzi et al. 2004).
These results
are still sensitive to Poissonian noise. Their
confirmation on a larger sample is a task
we are actively pursuing.

\begin{figure}
   \centering
   \includegraphics[width=9cm]{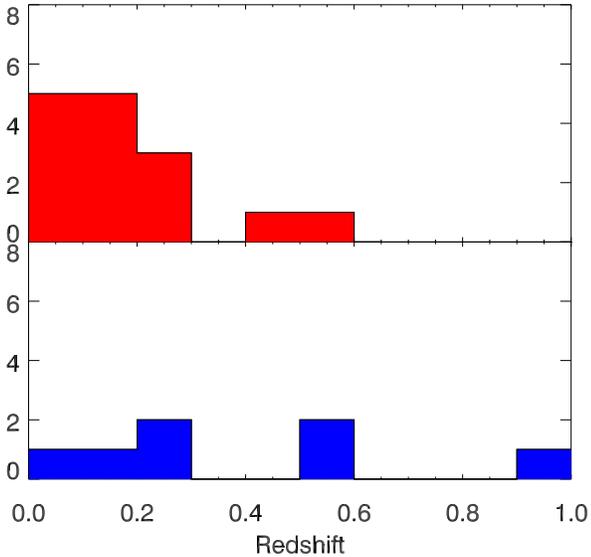}
\caption{Redshift distribution for the
GPS sample ({\it lower panel}) and the
``normal'' radio galaxy control sample
({\it upper panel})}
\label{fig8}
\end{figure}

Neither the control sample nor the GPS
sample are, by construction, unbiased or complete.
However, they are reasonably well matched in both
radio luminosity and redshift. The
K-S probability that
the radio luminosity distributions for GPS and
the control sample are drawn from the same parent
population is 66.5\%. The same probability for
the redshift distribution is somewhat smaller
(37.4\%; cf. Fig.~\ref{fig8}). This is potentially more worrying, as
the GPS redshift distribution median is
larger than in the control sample, and
obscured objects are in principle more easily
detected at higher redshifts. In order
to estimate the potential systematic error
due to comparing column density distributions corresponding
to samples with different redshift distributions,
we have simulated a large number of spectra, using
the best-fit models applicable to the GPS objects of
our sample, but assuming that they are
distributed in redshift in the same way as the objects of the
control sample. The difference between the
observed and the simulated GPS $N_H$ distributions
were parameterized through the
statistical uncertainty-weighted average of
the shift in the column densities,
$\Delta N_H$. The shifts are $-0.19 \pm 0.09$~dex
and $-0.08 \pm 0.09$~dex only, when the redshift
distributions for FR~II and FR~I are used.

Previous studies in the optical and UV
revealed a high detection rate for
FR~I RGs (\cite{chiaberge99}, 2002).
Their X-ray spectra are generally unobscured
(D04)
The emission of their central
compact core has been interpreted as
non-thermal synchrotron radiation from
the base of the jet, suggesting that the
we are observing the innermost regions close
to the supermassive black hole. In the framework
of the grand-unification scenario for FR sources
(\cite{urry95}) the larger obscuration of
GPS galaxies could be ascribed to their
orientation with respect to an
azimuthally-symmetric absorbing structure
(``torus'' hereafter),
which prevents us from directly observing
the central regions in objects observed
along lines-of-sight perpendicular to the
jet axis.

More puzzling is the evidence that GPS
galaxies are ``more obscured'' than FR~II
objects as well.
According to a classification originally due to
Jackson \& Rawlings (1997), FR~IIs can be
distinguished among Broad Line Objects (BLO), and
High- and Low-Excitation
galaxies (HEG, LEG; \cite{laing94}). BLOs
are interpreted as unobscured nuclei
(\cite{varano04}). Optical nuclei
of LEGs are detected in a comparable
fraction as FR~I, and are therefore
more similar to that class (which requires 
some revision of the standard zero-th order
unification scenario). Finally, 50\% of the HEG do
not show any optical counterpart, and should
represent the ``true'' torus-obscured radio-loud
AGN population. The covering fraction of
this ``torus'' seems to be redshift-dependent,
its opening angle widening up as the
accretion disk luminosity increases
(\cite{varano04}; see as well
Page et al. 2003; Jim\'enez-Bail\'on
et al. 2005). In our
control sample, both BLO (5 objects) and
LEG (2) FR~II RGs exhibit
typically $N_H \approxlt 3 \times
10^{21}$~cm$^{-2}$. The HEG (4)
show a much larger spread in column densities
($N_H \approxlt 10^{23}$~cm$^{-2}$),
with a
$\langle \log(N_{H,HEG}) \rangle = 2.7 \pm 0.6$,
consistent with that of the GPS galaxy sample
[$\langle \log(N_{H,HEG}) \rangle \ge 2.2 \pm 0.7$].
At face value, this implies that the line-of-sight to
HEG FR~II and GPS galaxies AGN
intercepts matter of comparable density, suggesting
a common orientation with respect to the obscuring
``torus'' postulated for the former.
No measurement of the orientation of the radio
structure is available for the objects of the
GPS sample. However, their symmetric morphology
suggests that the inclination angle is not small.
If X-rays originate close to the supermassive
black hole, and a torus surrounds the central region,
our line-of-sight will very likely intercept it.

Most of the GPS host galaxies
in our sample exhibit
a distorted morphology (4C+12.50,
PKS~0500+019, Mkn~668) and/or double nucleus
or close companions (all except COINS~J0029+3456,
\cite{peacock81,biretta85,fugmann88,stanghellini93,devries95,stickel96}).
High-resolution 2.2$\mu$ NICMOS/HST images
of 4C+12.50 and PKS~0941-080 confirm the
presence of double nuclei and, in the
former case, of a highly disturbed morphology
(Stanghellini et al., submitted; Guainazzi et al.,
in preparation; see as well de Vries et al. 1998, 2000
for ground-based near-infrared observations).
These reports indicate recent mergers.
The role of gravitational torques induced
by galaxy mergers in driving gas inflows toward the
nucleus has been highlighted by numerical simulations
(\cite{barnes96,taniguchi96}). Indeed, interacting
AGN pairs seem to invariably host X-ray obscured
nuclei (see Guainazzi et al. 2005, and references
therein). Galaxy mergers could therefore represent the
link between the large amount of gas in
the GPS nuclear environment and
the disturbed and multiple optical
morphologies of GPS host galaxies (\cite{odea96}).

Does the gas
probed by X-ray measurements have enough density
to influence the evolution of the parsec-scale
jet? Guainazzi et al. (2004) demonstrated
in the cases of Mkn~668 that if the parsec-scale
jet had to pierce its way through the Compton-thick
absorber covering the innermost nuclear
region, this could lead to an underestimate by
up to two orders of magnitude of the jet dynamical age
as derived from the hotspot recession velocity.
Even a Compton-thick absorber
would not be enough to ultimately ``choke'' the
expanding jet. However, measurements of
HI absorbing gas (\cite{pihlstrom03})
suggest a particle density
of $n_e \simeq$30~cm$^{-3}$ at the external
surface of the ``drilling jet'' (consistent
with a residual column density covering the soft X-ray
excess), which is still consistent with the ISM
density needed to ``frustrate'' its evolution.
For the GPS galaxies presented in this paper,
radio HI measurement are available only for PKS0500+019. The
$r \simeq$84~pc separated hot spots are covered by a
HI column density of $\simeq 6 \times 10^{20}$~cm$^{-2}$,
therefore about one order of magnitude lower
than observed in X-rays. This implies that the X-ray
absorber is located inside the hot-spots.
If the jet has to drill its way through it
before reaching the current separation,
the expansion time is:
$$
t_e \approxlt 7.3 \times 10^4 L^{-0.5}_{44} \Omega^{0.5}_{10} \hbox{years}
$$
(\cite{scheuer74,carvalho85})
where $L_{44}$ is the (unknown) luminosity injected in the jet
in units of $10^{44}$ and $\Omega_{10}$ is the jet
opening angle in units of 10$^{\circ}$. The
``braking effect'' of the X-ray absorber is
therefore negligible. Still,
any residual particle density facing the expanding
jet head is constrained only to be
$n_e \approxlt N_H/r = 20$~cm$^{-3}$. We
know very little of the distribution of the
ISM in the innermost kpc of this (and other) AGN and
the ultimate fate of the jet in PKS0500+019
is not known.

\begin{figure}
   \centering
   \includegraphics[width=9cm]{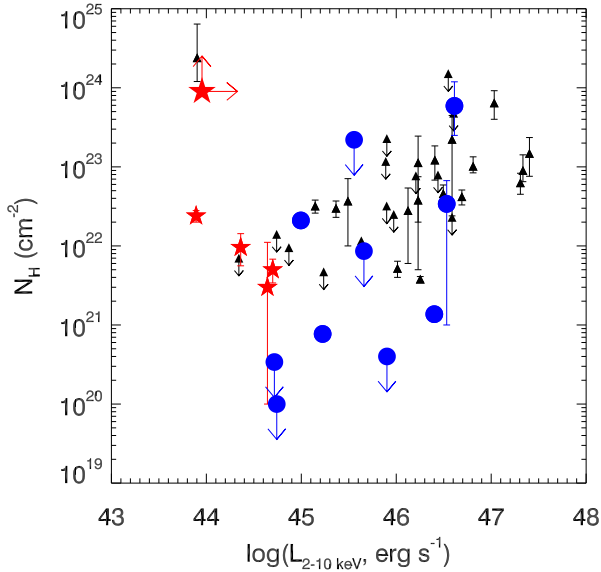}
\caption{$N_H$ versus X-ray luminosity
for GPS galaxies ({\it stars}) and
GPS/CSS quasars ({\it filled circles}).
The {\it small triangles} represent
the radio-loud quasars observed by
ASCA after Reeves \& Turner (2000)
              }
\label{fig9}
\end{figure}

With typical intrinsic X-ray luminosities
$\sim$10$^{44}$~erg~s$^{-1}$, most GPS galaxies of our
sample have a borderline luminosity
between the galaxy and the quasar regimes.
The X-ray spectral properties and luminosities
suggest that they may belong to the long-sought
class of obscured high-luminosity sources
predicted by theoretical models of the
Cosmic X-ray Background (CXB; \cite{gilli01}).
The cosmological evolution of GPS sources
- which represents a sizable fraction
of radio-selected extragalactic sources (\cite{odea98}) -
may have an impact on the CXB.
In Fig.~\ref{fig9}
we compare the $N_H$ versus
X-ray luminosity  in the GPS
galaxy sample with that measured in a sample of
GPS/CSS quasars (S99; \cite{siemiginowska03}; F05),
and in the large compilation of spectra of
radio-load quasars observed by ASCA after
Reeves \& Turner (2000). In the latter,
a correlation exists between the obscuring
column density and the absorption-corrected
2--10~keV luminosity. Given the large
fraction of upper limits on $N_H$ in
the sample, we have applied a linear 
fit $\log(N_H) = a + b \times \log(L_X)$
by extending the regression
method for left censored data described
by Schmitt (1985) and Isobe et al. (1986).
We performed a large number of least square
fits on a set of Monte-Carlo simulated
data derived from the experimental
points according to the following rules:
a) each detection was substituted by a random
Gaussian distribution, whose
mean is the best-fit measurement and whose
standard deviation is its statistical uncertainty;
b) each upper limit $U$ was substituted by a random
uniform distribution in the
close interval [0,U]. 
The slope obtained with this
method - albeit rather shallow -
is larger than 0 at a confidence
level greater than 5$\sigma$:
$b_{QSO} = (3.9 \pm 0.7) \times 10^{-3}$.
The same fit applied on the GPS/CSS
sample yields no significant evidence for
a correlation, even if we include the
GPS {\it galaxies} discussed in this paper:
$b_{GPS/CSS} = (-30 \pm 100) \times 10^{-3}$,
although the large statistical uncertainties prevent
us from drawing any firm conclusions.
If this different behavior is
confirmed on a larger sample,
this may indicate that GPS/CSS quasars are a ``mixed
bag'' of obscured/unbeamed and unobscured/beamed
objects. 
Beamed structures in high-luminosity quasars
may create an unobscured line-of-sight by
piercing their way
through dense nuclear environment gas.
In statistical terms, this would happen more rarely
in lower-luminosity
galaxies. This scenario cannot be investigated
with the currently available sample, and would
require good quality X-ray observations on
a significantly larger, statistically unbiased
sample of GPS sources.

\section{Conclusions}

This paper investigates the apparent
X-ray under-luminosity of GPS sources
shown by ROSAT (\cite{baker95}), by
obtaining for the first time good quality X-ray
spectra on a sizable sample thereof.
The main conclusions of
this study are:

\begin{itemize}

\item The GPS
radio-to-X-ray Spectral Energy distributions are
generally similar.
The ratio between the 5~GHz and the 5~keV flux density is
constrained within a factor $\simeq$15.
There is no unique reason for this residual
scatter, which is likely to be due to
a combination of X-ray obscuration and
intrinsic luminosity effects. On the
other hand, PKS~0941-080
is two orders of magnitude X-ray under-luminous
with respect to the average of the
other objects discussed in this
paper. Whether this is due to high obscuration or
intrinsic X-ray weakness is impossible to tell
with the current data. We therefore refrain 
from inferring any implications from this single
outlier

\item The detection of rapid 
variability during the XMM-Newton observation of
COINS~J0029+3456 puts a constraint on the
high-energy photon emitting region $\approxlt$10~$\mu$pc,
supporting an origin at the base of the expanding jet
(where radio synchrotron emission is likely to
be self-absorbed) or in
the accretion disk surrounding the active nucleus.

\item The obscuring screen in PKS0500+019
is located well within the radio hot spots,
as suggested by the fact that the column density
determined in X-rays is almost one order of magnitude
higher than that estimated from radio measurements. The
hot spots are therefore completely X-ray silent
in PKS0500+019. The lack of soft excess emission
above the obscured X-ray emission in COINS~J0029+3456
can be interpreted likewise

\item GPS galaxies exhibit a radio-to-X-ray
luminosity ratio comparable to a control sample of ``normal'' radio
galaxies. This indicates that it is unlikely that GPS galaxies
are {\it intrinsically} X-ray weaker than
their less compact counterparts

\item The X-ray obscuring gas column density distribution
function for GPS galaxies is inconsistent at the
$\approxgt$98.7\% level
with the distribution function of a control sample of
RGs. The main reason for this difference is
the comparative paucity of X-ray unobscured
($N_H \le 10^{21.5}$~cm$^{-2}$) GPS galaxies

\item The average X-ray column density in our sample is
consistent with that observed in the 4 HEG FR~II
galaxies for which measurements of
this observable are available. This suggests that
X-ray absorption in a GPS galaxy is mainly due to
orientation with respect to an obscuring ``torus''

\item There is no compelling evidence for 
hot ($kT \sim 10^6$-$10^7$~K) gas from the X-ray spectra

\end{itemize}

The above
evidence support a scenario whereby GPS radio galaxies
are ``young'' counterparts of large-scale FR~II
radio galaxies.
$N_H \sim 10^{22}$~cm$^{-2}$
column densities in the innermost kpc around the
active nucleus are unlikely to substantially affect
the evolution of sub-kpc jet structures. However,
tenuous ISM matter could have a much more important
role in braking or smothering the full expansion of the
radio structure. Signatures of this hot gas could
be revealed by deep high-resolution X-ray {\it Chandra}
observations.

Although the GPS sample presented in this paper represents
the largest compilation ever published of X-ray measurements of
this class of objects,  the sample
is small and
admittedly far from complete or unbiased. On the
contrary, it is clearly biased toward X-ray brightest (and/or
therefore potentially less obscured) objects. Moreover, an accurate
determination of
the ionization state of the absorber is beyond the capability
of the current high-resolution X-ray detectors for objects
of X-ray flux $\sim 10^{-13}$~erg~cm$^{-1}$~s$^{-1}$,
and must await the next generation of X-ray missions.
Questions such as the nature of the gaseous nuclear
environment of the early evolutionary
stages of AGN (\cite{dimatteo05}), or the
feedback between the (recurrent? \cite{siemiginowska02}, 2003)
AGN activity and its environment, or surrounding clusters
(\cite{fabian03, siemiginowska05a}) could be properly addressed
by high quality spectroscopic
X-ray measurements of a sizable sample of GPS
sources.
Nonetheless,
the results presented in this paper open a so far unexplored window on
the environmental conditions surrounding the origin of the
radio power in the Universe. An extension of this study
to a complete and unbiased sample of GPS galaxies is under way.

\begin{acknowledgements}
This paper is based on observations obtained with XMM-Newton, an ESA
science mission with instruments and contributions directly funded by
ESA Member States and the USA (NASA).
This research has made use of
data obtained through the High Energy Astrophysics Science Archive
Research Center Online Service, provided by the NASA/Goddard Space
Flight Center and of the NASA/IPAC Extragalactic Database (NED) which
is operated by the Jet Propulsion Laboratory, California Institute of
Technology, under contract with the National Aeronautics and Space
Administration. This work was partly supported by NASA grants GO2-3148A and
NNG04GF98G. AS acknowledges support by NAS8-39073. Careful reading
of the manuscript by
the anonymous referee is gratefully acknowledged.

\end{acknowledgements}


\begin{thebibliography}{}

\bibitem[Baker et al. 1995]{baker95} Baker J.C., Hunstead R.W., Brinkmann W., 1995, MNRAS, 277. 553

\bibitem[Barnes \& Hernquist 1996]{barnes96} Barnes J.E., Hernquist L., 1996, ApJ, 471, 115

\bibitem[Bennett et al. 2003]{bennett03} Bennett C.L., et al., 2003, ApJS, 148, 1

\bibitem[Baum et al. 1990]{baum90} Baum S.A., O'Dea C.P., de Bruyn A.G., Murphy D.W., 1990, A\&A, 232, 19

\bibitem[Biretta et al. 1985]{biretta85} Biretta J.A., Schneider D.P., Gunn J.E., 1985, AJ, 90, 2508

\bibitem[Carilli et al. 1998]{carilli98} Carilli C.L., Menten K.M., Reid M.J., Rupen M.P., Yun M.S., 1998, ApJ, 494, 175

\bibitem[Carvalho 1985]{carvalho85} Carvalho J.C., 1985, A\&A, 150, 129

\bibitem[Carvalho 1994]{carvalho94} Carvalho J.C., 1994, A\&A, 292, 392

\bibitem[Carvalho 1998]{carvalho98} Carvalho J.C., 1998, A\&A, 329, 845

\bibitem[Chiaberge et al. 1999]{chiaberge99} Chiaberge M., Capetti A., Celotti A., 1999, A\&A, 349, 77

\bibitem[Chiaberge et al. 2002]{chiaberge02} Chiaberge M., Duccio Macchetto F., Sparks W.B., et al., 2002, ApJ, 571, 247

\bibitem[Condon et al. 1998]{condon98} Condon J.J., Cotton W.D., Greisen E.W., et al., 1998, AJ, 115, 1693

\bibitem[Dallacasa et al. 1997]{dallacasa97} Dallacasa D., Bondi M., Alef W., Mantovani F.,
1997, A\&A, 325, 943

\bibitem[de Vries et al. 1995]{devries95} de Vries W.H., Barthel P.D., Hes R., 1985, A\&AS, 114, 259

\bibitem[de Vries et al. 1998]{devries98} de Vries W.H., O'Dea C.P., Perlman E., et el., 1998, ApJ, 503, 138

\bibitem[de Vries et al. 2000]{devries00} de Vries W.H., O'Dea C.P., Barthel P.D., Thompson D.J., 2000, A\&AS, 143, 181

\bibitem[De Young 1993]{deyoung93} De Young D.S., 1993, ApJ, 402, 95

\bibitem[Dickey \& Lockman 1990]{dickey90} Dickey J.M., Lockman F.J., 1990, ARA\&A 28, 215

\bibitem[di Matteo et al. 2005]{dimatteo05} di Matteo T., Springel V., Hernquist L., 2005, Nature, 433, 604

\bibitem[Donato et al. 2004]{donato04} Donato D., Sambruna R.M., Gliozzi M., 2004, ApJ, 617, 915 (D04)

\bibitem[Edwards \& Tringay 2004]{edwards04} Edwards P.G., Tringay S.J., 2004, A\&A, 424, 91

\bibitem[Fabian et al. 2003]{fabian03} Fabian A.C., Sanders J.S., Allen S.W., et al., 2003, MNRAS, 344, L43

\bibitem[Fanti et al. 1995]{fanti95} Fanti C., Fanti R., Dallacasa D., et al., 1995, A\&A, 302, 317

\bibitem[Fiocchi et al. 2005]{fiocchi05} Fiocchi M.T., et al., 2005, ApJ, submitted (F05)

\bibitem[Fossati et al. 1998]{fossati98} Fossati G., Maraschi L., Celotti A., Comastri A., Ghisellini G., 1998, MNRAS 299, 433

\bibitem[Fugmann \& Meisenheimer 1988]{fugmann88} Fugmann W., Meisenheimer K., 1988, A\&AS, 76, 145

\bibitem[Gabriel et al. 2003]{gabriel03} Gabriel C., Denby M., Fyfe D. J., Hoar J., Ibarra A., 2003, in ASP Conf. Ser., Vol. 314 Astronomical Data Analysis Software and Systems XIII, eds. F.Ochsenbein, M.Allen, D.Egret (San Francisco: ASP), 759 

\bibitem[Gehrels 1986]{gehrels86} Gehrels N,, 1986, ApJ, 303, 336

\bibitem[Gilli et al. 2001]{gilli01} Gilli R., Salvati M., Hasinger G., 2001, A\&A, 366, 407

\bibitem[Gopal-Krishna \& Wiita 1991]{gopalkrishna91} Gopal-Krishna, Wiita P.J., 1991, ApJ, 373, 325

\bibitem[Guainazzi et al. 2000]{guainazzi00} Guainazzi M., Oosterbroek T., Antonelli L.A., Matt G., 2000, A\&A, 364, L80

\bibitem[Guainazzi et al. 2005]{guainazzi05} Guainazzi M., Piconcelli E., Jimenez-Bil\'on E., Matt G., 2005, A\&A, 429, L9

\bibitem[Guainazzi et al. 2004]{guainazzi04} Guainazzi M., Siemiginowska A., Rodriguez-Pascual P., Stanghellini C., 2004a, A\&A, 421, 461

\bibitem[Isobe et al. 1986]{isobe86} Isobe T., Feigelson E.D., Nelson P.I., 1986, ApJ, 306, 490

\bibitem[Iwasawa et al. 1999]{iwasawa99} Iwasawa K., Allen S., Fabian A.C., Edge A.C., Ettori S., 1999, MNRAS, 306, 467

\bibitem[Jackson \& Rawlings 1997]{jackson97} Jackson N., Rawlings S., 1997, MNRAS, 286, 241

\bibitem[Jim\'enez-Bail\'on et al. 2005]{jimenezbailon05} Jim\'enez-Bail\'on E., Piconcelli E., Guainazzi M., et al.,
2005, A\&A, 435, 449

\bibitem[Kadler et al. 2004]{kadler04} Kadler M., Kerp. J., Ros E., et al., 2004, A\&A, 420, 467

\bibitem[Kollgaard et al. 1995]{kollgaard95} Kollgaard R.I., Feigelson E.D., Laurent-Muehleisen S.A., et al., 1995, ApJ, 449, 61

\bibitem[K\"uhr et al. 1981]{kuhr81} K\"uhr H., Witzel A., Pauliny-Toth I.I.K., Nauber U., 1981, A\&AS, 45, 367

\bibitem[Laing et al. 1994]{laing94} Laing R.A., Jenkins C.R., Wall J.V., Unger S.W., in ``The First Stromlo Symposium: The Physics of Active Galactic Nuclei'', Bicknell G.V., Dopita M.A., Quinn P.A., ASP Conf. Ser., 54, 201

\bibitem[Ma et al. 1998]{ma98} Ma C., Arias E.F., Eubanks T.M., et al., 1998, AJ, 116, 516

\bibitem[Mazzarella et al. 1991]{mazzarella91} Mazzarella J.M., Bothun G.D., Boroson T.A., 1991, AJ, 101, 2034

\bibitem[Murgia 2003]{murgia03} Murgia M., 2003, PASA, 20, 19

\bibitem[O'Dea et al. 1998]{odea98} O'Dea C., PASP, 107, 803

\bibitem[O'Dea et al. 1996]{odea96} O'Dea C., Stanghellini C., Baum S., Charlot S., 1996, ApJ, 470, 806

\bibitem[O'Dea et al. 2000]{odea00} O'Dea C., de Vries W.H., Worrall D.M., Baum S., Koekmoer A., 2000, AJ, 119, 478

\bibitem[Page et al. 2003]{page03} Page K.L., O'Brian, Reeves J.N., Breeveld A.A., 2003, MNRAS, 340, 1052

\bibitem[Peacock et al. 1981]{peacock81} Peacock J.A., Perryman M.A.C., Longair M.S., Gunn J.E., Westphal J.A., 1981, MNRAS, 194, 601

\bibitem[Perola et al. 2002]{perola02} Perola G.C., Matt G., Cappi M.,
et al., 2002, A\&A, 389, 202

\bibitem[Phillips \& Mutel 1982]{phillips82} Phillips R.B., Mutel R.L., 1982, MNRAS, 257, L19

\bibitem[Pihlstr\"om et al. 2003]{pihlstrom03} Pihlstr\"om Y.M., Conway J.E., Vermeulen R.C., 2003, A\&A, 404, 871

\bibitem[Polatidis \& Conway 2003]{polatidis03} Polatidis A.G., Conway J.E., 2003, PASA, 20, 69

\bibitem[Readhead et al. 1996]{readhead96} Readhead A.C.S., Taylor G.B., Pearson T.J., Wilkinson P.N., 1996, ApJ, 460, 634

\bibitem[Reeves \& Turner 2000]{reeves00} Reeves J.N., Turner M.J.L.,
2000, MNRAS, 316, 234

\bibitem[Reeves et al. 1997]{reeves97} Reeves J., Turner M.J.L., Kii T., Ohashi T., 1997, MNRAS, 312, L17

\bibitem[Risaliti 2002]{risaliti02} Risaliti G., 2002, A\&A, 386, 379

\bibitem[Sambruna et al. 1999]{sambruna99} Sambruna R., Eracleous M., Mushotzky R., 1999, ApJ, 526, 60 (S99)

\bibitem[Satypal et al. 2004]{satyapal04} Satypal S., Sambruna R.M., Dudik R.P., 2004, A\&A, 414, 8235

\bibitem[Scheuer 1974]{scheuer74} Scheuer P.A.G., 1974, MNRAS, 166, 513

\bibitem[Schmitt 1985]{schmitt85} Schmitt J.H.M.M., 1985, A\&A, 293, 178

\bibitem[Siemiginowska et al. 2002]{siemiginowska02} Siemiginowska A., 
Bechtold J., Aldcroft T.L., et al., ApJ, 570, 543

\bibitem[Siemiginowska et al. 2003]{siemiginowska03} Siemiginowska A., Aldcroft T.L., Bechtold J., et al., 2003, PASA, 20, 113

\bibitem[Siemiginowsha et al. 2005a]{siemiginowska05a} Siemiginowska A., Cheung C.C., LaMassa S., et al., 2005a, ApJ, submitted

\bibitem[Siemiginowska et al. 2005b]{siemiginowska05b} Siemiginowska A., et al., 2005b, ApJ, submitted

\bibitem[Snellen et al. 1996]{snellen96} Snellen I.A.G., Bremer M.N., Schillizzi R.T., et al., 1996, MNRAS, 279, 1294

\bibitem[Stanghellini et al. 2001]{stanghellini01} Stanghellini C.,
Dallacasa D., O'Dea C.P., et al., 2001, A\&A, 377, 377
(erratum: 379, 870)

\bibitem[Stanghellini et al. 1993]{stanghellini93} Stanghellini C.,
O'Dea C.P., Baum S.A., Laurikainen E., 1993, ApJS, 88, 1

\bibitem[Stanghellini et al. 1997]{stanghellini97} Stanghellini C.,
O'Dea C.P., Baum S.A., et al., 1997, A\&A, 325, 953

\bibitem[Stanghellini et al. 1998]{stanghellini98} Stanghellini C.,
O'Dea C.P., Dallacasa D., et al., 1998, A\&AS, 131, 303

\bibitem[Stickel et al. 1996]{stickel96} Stickel M., Rieke M.J., Rieke G.H., Kuehr H., 1996, A\&A, 306, 49

\bibitem[Str\"uder et al. 2001]{struder01} Str\"uder L., Briel U., Dannerl K., et al., 2001, A\&A 365, L18

\bibitem[Taniguchi \& Wada 1996]{taniguchi96} Taniguchi Y., Wada K., 1996, ApJ, 469, 581

\bibitem[Tingay et al. 2003]{tingay03} Tingay S.J., Edwards P.G., Tzioumis A.K., 2003, MNRAS, 346, 327

\bibitem[Turner et al. 2001]{turner01} Turner M.J.L., Abbey A., Arnaud M., et al., 2001, A\&A 365, L27

\bibitem[Urry \& Padovani 1995]{urry95} Urry C.M., Padovani P., 1995, PASP, 107, 803

\bibitem[Varano et al. 2004]{varano04} Varano S., Chiaberge M., Macchetto F.D., Capetti A., 2004, A\&A, 428, 401

\bibitem[Voges et al. 2000]{voges00} Voges W., Aschenbach B., Boller Th., et al., 2000, IAUC 7432

\bibitem[Weisskopf et al. 2002]{weisskopf02} Weisskopf M.C., Brinkman B., Canizares C., et al., 2002, PASP, 114, 1

\bibitem[Zensus et al. 2002]{zensus02} Zensus J.A., Ros E., Kellermann K.I., et al., 2002,  AJ, 124, 662

\end{thebibliography}
\end{document}